\documentclass[12pt,preprint]{aastex}

\slugcomment{To appear in ApJ main Journal}

\shorttitle{massive nearby cooling core cluster}
\shortauthors{B\"ohringer et al.}

\begin{document}

\title{CHANDRA reveals galaxy cluster with the most massive nearby
  cooling core, RXCJ1504.1-0248}

\author{H. B\"ohringer, V. Burwitz
 Y.-Y. Zhang, P. Schuecker, and N. Nowak}
\affil{Max-Planck-Institut f\"ur extraterr. Physik,
                 D 85741 Garching, Germany}


\begin{abstract}
A CHANDRA follow-up observation of an X-ray luminous galaxy cluster with a
compact appearance, RXCJ1504.1-0248
discovered in our REFLEX Cluster Survey, reveals
an object with one of the most prominent cluster cooling cores. A
$\beta$-model fit to the X-ray surface brightness profile gives a core
radius of $\sim 30 h_{70}^{-1}$ kpc which is much smaller than the
cooling radius with $\sim 140$ kpc. As a consequence more than 70\% of
the high X-ray luminosity of $L_{bol} = 4.3~ 10^{45} h_{70}^{-1}$ erg s$^{-1}$
of this cluster is radiated inside the cooling radius. A simple
modeling of the X-ray morphology of the cluster leads to a formal mass
deposition rate within the classical cooling flow model of $1500 -
1900$ M$_{\odot}$ yr$^{-1}$ (for $h_{100}=0.7$, and $2300 - 3000$ 
M$_{\odot}$ yr$^{-1}$ for $h_{100}=0.5$).

The center of the cluster is marked by a
giant elliptical galaxy which is also a known radio source. Thus it is
very likely that we observe one of the interaction systems where the
central cluster AGN is heating the cooling core region in a
self-regulated way to prevent a massive cooling of the gas, similar to
several such cases studied in detail in more nearby clusters. The
interest raised by this system is then due to the high power
recycled in RXCJ1504-0248 over cooling time scales
which is about one order of magnitude higher
than what occurs in the studied, nearby cooling core clusters.
The assumption that cooling is exactly balanced by
the AGN heating implies a central black hole mass growth rate of
the order of 0.5 M$_{\odot}$ yr$^{-1}$.
This cluster is therefore a prime target for the study of 
AGN-intracluster medium interaction at very extreme conditions.
Further features common to cooling cores found in this cluster are
a strong temperature drop towards the center and narrow, low
ionization emission lines in the central cluster galaxy.

The cluster is also found to be very massive, with a global X-ray
temperature of about 10.5 keV and a total mass 
of about $1.7~10^{15} h_{70}^{-1}$ M$_{\odot}$
inside  $3 h_{70}^{-1}$ Mpc.
\end{abstract}


\section{Introduction}

One of the currently most debated questions concerning the structure
of the X-ray luminous, hot intracluster plasma of clusters of galaxies
is the consequence of the small cooling time of this plasma
in those clusters with dense central cores (e.g. Fabian 1994). 
XMM-Newton X-ray spectroscopy has shown that in spite
of its short cooling time the gas is not cooling at the high expected
rates in the absence of heating processes (e.g. Peterson 2001, 2003;
Matsushita et al. 2002; B\"ohringer et al. 2002; Molendi 2002). The most popular
scenario which allows for a self-regulated heating of the hot plasma
in cluster cooling cores and prevents massive cooling is the heating
of the intracluster medium (ICM) by the jets and radio lobes of the AGN
in the central cluster galaxies (e.g. Churazov et al. 2000, 2001; McNamara et al. 2000;
David et al. 2001; Fabian 2003; Forman et al. 2004). It was shown
that this interaction provides enough power in many nearby cooling flows
to at least balance cooling. For example the current kinetic energy output
of the inner radio lobes in M87 in Virgo and NGC 1275 in the Perseus
cluster is with estimated values of $\sim 10^{44}$ erg s$^{-1}$ and
$\sim 10^{45}$ erg s$^{-1}$, respectively, in both these cases about an order of
magnitude higher than the cooling power in the cooling core 
(Churazov et al. 2000, 2003; Birzan et al. 2004). 
Recently further very deep and detailed X-ray observations have
given some insight into the details of the heating process, which is
proposed to occur through the dissipation of sound waves set off by
the interaction of the radio lobes with the ICM \citep{fab03}
or by shock fronts (Forman et al. 2004, 
Nulsen et al. 2005a, 2005b, McNamara et al. 2005). A promising picture seems
to be emerging where the recycling of AGN energy with powers in the order of
$10^{44}$ to $10^{45}$ erg s$^{-1}$ in cluster cooling cores
of nearby clusters could explain the observed phenomena.

Searching through larger volumes of our Universe, more extreme cases of cluster
cooling cores can be found, where the rate of AGN energy recycling is even higher
by up to an order of magnitude. The cluster with the largest cooling core
detected so far, RXCJ1347.5-1144 at z=0.4516
was discovered in the REFLEX survey \citep{schin95,boe04}
with a formally derived cooling flow mass deposition rate
of the order of 3000 M$_{\odot}$ yr$^{-1}$ (for a Hubble constant of
$H_0 = 50$ km s$^{-1}$ Mpc$^{-1}$; Schindler et al. 1997; Allen 2000)
which corresponds to an energy dissipation
of $\sim 10^{46}$ erg s$^{-1}$. This cluster is unfortunately much more distant
than the well studied nearby clusters and does not lend itself easily
to a detailed observational study.
We have now discovered a more nearby, similarly striking massive
cooling core cluster in the REFLEX Survey, RXCJ1504.1-0248 at a redshift
of $z=0.2153$. This cluster was flagged as a cluster candidate due to
a galaxy overdensity detected in the COSMOS data base \citep{cos84} 
and six concordant galaxy redshifts found in our subsequent
follow-up observations confirmed the existence of a cluster.
Three of these cluster galaxies show AGN-like spectra.

Serious doubts remained about the cluster identification of this X-ray emitter,
because the X-ray source appeared much too compact for its high luminosity,
compared to the other clusters in the REFLEX sample in the same
distance and luminosity range. This could possibly be attributed to a
contaminating central AGN. The certain identification clearly required a
higher resolution X-ray observation, which could recently be made
through a CHANDRA snap-shot exposure yielding a perfect cluster image without
significant contamination by point sources (Fig.1). With these source properties
it becomes immediately clear that RXCJ1504.1-0248 must have an extremely bright core
and is potentially a very interesting cooling core cluster.

In this paper we study the structure of this cluster in more detail.
In section 2 and 3 we present the observational results. Section
4 is devoted to the modeling of the mass profile and the determination
of the parameters in the frame of a classical cooling flow model.
In section 5 we discuss further phenomenological features related to
the cooling core of the cluster and section 6 provides the conclusion.
We will adopt a cosmological model with $\Omega_m = 0.3$,
$\Omega_{\Lambda} = 0.7$ and $H_0 = 70$ km s$^{-1}$ Mpc$^{-1}$, except
if it is noted otherwise. Thus 1 arcmin at the distance of RXCJ1504-0248
corresponds to 209 kpc.

\section{Observation and Data preparation}

RXCJ1504.1-0248 was observed with the CHANDRA ACIS-I on January 7, 2004 for
13463 sec. The observation was hardly disturbed by times of high background
and the net exposure after standard cleaning procedures is 13 298 sec.
Fig. 2 shows an image of the cluster in the 0.5 - 2 keV energy band
superposed on an optical R-band image taken as 5 min exposure in 
our REFLEX redshift survey at the ESO La Silla 3.6m telescope.
The total ACIS-I count rate in the region $r \le 6$ arcmin in the 0.5
to 2 kev band is 2.03 cts s$^{-1}$ implying an unabsorbed flux of about
$1.18$ and $1.9 \cdot 10^{-11}$ erg s$^{-1}$ cm$^{-2}$ in the $0.5 -
2$ kev and $0.1 - 2.4$ keV energy bands, respectively. In the ROSAT
All-Sky Survey we found a flux of $2.2 (\pm 0.11)  \cdot 10^{-11}$ erg s$^{-1}$
cm$^{-2}$ in the $0.1 - 2.4$ keV band within an aperture of
12 arcmin radius, in reasonable agreement with the new results from
the much deeper image. 

The X-ray cluster center coincides well with a central dominant galaxy
that can be identified with LCRS B150131.5-023636 at the J2000 position 
15 04 07.5 -02 48 16 at z=0.216917 (Shechtman et al. 1996). The optical image of
the cluster and the spectrum of the galaxy B150131.5-023636 described in section 5
have been obtained with EFOSC2 at the 3.6m telescope of ESO La Silla on Aug. 14
and 20, 2001, respectively.

\section {Analysis and Results}

The above noted fluxes imply an X-ray luminosity of $L_x = 2.3 \cdot 10^{45}
h_{70}^{-1}$ erg s$^{-1}$ in the $0.1 - 2.4$ keV band and a bolometric
X-ray luminosity of $L_x = 4.3 \cdot 10^{45}h_{70}^{-1}$ erg s$^{-1}$.
This makes this cluster the most prominent X-ray luminous cluster in
the southern sky at redshifts below $z = 0.34$, with only two galaxy
clusters at larger distances in the REFLEX catalogue having a higher X-ray luminosity.
The X-ray image in Fig. 1 shows a high degree of regularity with
a slightly elliptical shape and a major axis approximately along a 
position angle of about 40 degrees (North to East). 
As seen in Fig. 2 the center of the X-ray emission 
is marked by a dominant giant galaxy in the optical.

The azimuthally averaged X-ray surface brightness profile from which the
background was subtracted is well described by a
$\beta$-model out to a radius of about 300 arcsec, outside which the
background subtraction uncertainties become significant (Fig. 3).
The background was estimated either from the outer region of
the CCD or from an external background field. Remarkable
is the small core radius of $r_c \sim 30~  h_{70}^{-1}$ kpc 
and the high central gas density of $n_{e0} \sim 0.16 h_{70}^{1/2}$ 
cm$^{-3}$. The slope parameter, $\beta$, has the very typical value of $0.6$.  

The temperature profile was determined by fitting a spectral MEKAL model 
with fixed galactic absorption with a column density of $6~ 10^{20}$
cm$^{-2}$ (as measured from 21 cm observations, Dickey \& Lockman 1990)
to the spectra obtained from the photons extracted from concentric rings around
the cluster center. The background for subtraction was taken either
from a background region at the outer parts of the detector (radial zone
3.8 to 5.7 arcmin) or from a background field in the same rings as taken for the
target spectral extraction. Since there is some faint X-ray emission from the cluster
almost throughout the entire detector, the on-target background subtraction 
prevents an accurate temperature determination at larger radii. Fig 4
provides a comparison of both methods, showing that the two approaches yield 
practically identical results out to a radius of 1 arcmin, but the analysis
based on an external background field can be extended to a radius of 3
arcmin. While the bulk temperature of the cluster is about 10.5 keV, we note a
strong temperature drop towards the center to a value below 5 keV. Such
a temperature drop by a factor of 2 or 3 is observed in many cooling 
core clusters (e.g. De Grandi \& Molendi 2002; Fabian 2003; Ikebe et al. 2004;
Sanders et al. 2004). 
A possible temperature drop to larger radii indicated by the data
cannot be established with the present photon statistics. 
Fig. 5 shows the spectrum of the innermost circle (0 - 15 arcsec). For the
given photon statistics the spectrum is well fit by a one-temperature MEKAL
model. We do not note any features which could indicate the Fe L line
complex observed at lower temperatures, which would indicate the presence of
cooler temperature phases.  

In Fig. 4 we show 3 rough fits of analytic expressions to the temperature
profile which were chosen to approximately bracket the inner and outer gradients of
the temperature profile. The analytical expressions are:

$T_1 ({\rm keV}) = 3.5 + 0.44\cdot r^{0.9} - 0.044 \cdot r^{1.4} + 0.0007 \cdot
r^{2}$ (for $r<150$)

\hskip 2.5cm and $T_1 = 11.076 - 0.00544    \cdot r$ (for $r >150$)

$T_2 ({\rm keV}) = 3.0 + 0.305\cdot r^{0.9} - 0.0192\cdot r^{1.4}$ 
(for $r<100$)

\hskip 2.5cm and $T_2 = 9.80 + 0.003322 \cdot r$ (for $r > 100$)

$T_3 ({\rm keV}) = 3.5 + 0.52\cdot r^{0.9} - 0.0526 \cdot r^{1.4} + 0.00083\cdot
r^2$ (for $r<150$)

\hskip 2.5cm and $T_3 = 12.932 - 0.01364 \cdot r$ (for $r > 150$) 

\noindent
where the radius, r, is in units of arcsec.

The spectral fits also indicate abundances of heavy elements (dominated by the
fit to the Fe K line) of about 0.3 to 0.4 solar with large errors of 
about 0.1 to 0.2 in solar units
(abundances based on Anders \& Grevesse 1989). To our surprise we do not find 
a strong increase of the iron abundance towards the center, as seen in many
cooling flow clusters (DeGrandi \& Molendi 2001), but a deeper 
observation is necessary to draw a firm conclusion. 

\section{Modeling the cluster structure}

\subsection{Mass profile}

From an analytical deprojection of the $\beta$-model fit to the X-ray surface
brightness profile of the cluster and the temperature profile we can obtain
the cluster mass profile under the assumption of hydrostatic equilibrium and
spherical symmetry. The resulting mass profile is shown in Fig. 6 together
with the profile of the gas mass, as derived from the $\beta$-model fit. For
the gravitational mass profile we also indicate the typical uncertainties
determined from the local minima and maxima of the
mass profile for the different analytical temperature fits shown in Fig. 4. 
An additional scaling uncertainty of 15\% for $T_1$ and $T_2$ and
5\% for $T_3$ is included. At a radius of $3~h_{70}^{-1}$ Mpc the total mass is 
$1.8 (+0.35, -0.28)~10^{15}~h_{70}^{-1}$~M$_{\odot}$. For the radius $r_{200} =
2.3$ Mpc we find a mass of $1.5 ~10^{15}~h_{70}^{-1}$~M$_{\odot}$. Thus
RXCJ1504-0248 is among the most massive clusters known. The gas-to-total mass
ratio for the two fiducial radii is $0.17 (+0.03,-0.07)$ and 
$0.15 (+0.03,-0.05)$, respectively.

\subsection{Cooling flow analysis}

To gain an understanding of the processes occuring in the central 
ICM of this cluster we start with a classical cooling flow analysis 
(e.g. Fabian et al. 1984; Thomas et al. 1987). We take two approaches. For
the more simple model A we equate the energy loss by radiation 
inside a given radius, $r$, with the enthalpy influx from outside 
through the sphere with radius $r$. In model B we formulate the energy balance
in a local differential way and include the gain of gravitational energy of
the material flowing in from the outer regions.

$$ \dot M(r) = 4 \pi r^2  {n_e^2 \Lambda(T) + {\dot M(r+\Delta r) \over
    \Delta r 4\pi r^2} \cdot {5k_B T \over 2 \mu m_p} \over  
{5k_B \over 2 \mu m_p} \left( {T \over \Delta r} + {dT \over dr}\right) 
+ {M(r)G \over r} + {\Phi \over \Delta r} } $$

\noindent
where $\Lambda(T)$ is the cooling function normalized to the electron density
squared, $\Delta r$ is the shell width in the numerical calculation, $\Phi$
is the gravitational potential, and the other symbols have their usual meaning.

Fig. 7 shows the cooling time as a function of the cluster radius. If we take
the often used fiducial value of $10^{10}$ years, we find a cooling radius of 
$140 (\pm 5)$ kpc (for a Hubble constant of 
$h_{100} = H_0 /(100$ km s$^{-1}$ Mpc$^{-1}) = 0.7$)
and $165 (\pm 5)$ kpc (for $h_{100} = 0.5$). Here the
uncertainty is determined from the minima and maxima for the different adopted
fits to the temperature profile. Fig. 8 then shows the mass flow rates
determined for model A and B and the cooling radius is indicated by vertical
lines. For this adopted cooling radius we find mass flow rates of
1400 and 1900 M$_{\odot}$ yr$^{-1}$ (for $h_{100} = 0.7$) and 2300 and 2930
(for $h_{100} = 0.5$) for model B and A, respectively. 

These high formal cooling flow mass deposition rates make RXCJ1504-0248 the most
prominent cooling flow cluster next to the most luminous cluster known,
RXCJ1347-1144 at $z = 0.45$  for which also a mass deposition rate of the order of 3000  
M$_{\odot}$ yr$^{-1}$ (for $h_{100} = 0.5$) was deduced (Schindler et
al. 1997; Allen 2000). This makes RXCJ1504-0248 a very interesting
target to study the cooling core phenomenon under the most extreme conditions.

\section{Discussion}

In the new scenario of the physics of cooling core clusters, the large
radiative cooling rates are compensated by the energy released from a central
AGN (e.g. Churazov et al. 2000, 2001; McNamara et al. 2000; David et al. 2001;
B\"ohringer et al. 2002; Fabian 2003; Forman et al. 2004). To keep the
balance such that neither massive mass condensation nor a dispersion of the dense
gaseous core occurs, the heating has to be fine-tuned. This is achieved by a
self-regulation system where large mass deposition rates in the center lead to
an increased feedback from the AGN which limits the cooling rate. 
Seen from the perspective of the central AGN, its accretion rate is limited by
the amount of cooling that can occur, that is the energy that can be
dissipated by the ICM in the cooling core region (Churazov et al. 2002). 
In this cooling core scenario a high radiative power of the central ICM
indicates a fast accretion of the interacting AGN. 

The case of RXCJ1504-0248 is an extreme case in this scenario. Since the core
radius is so small, actually much smaller than the cooling radius by a factor
of about four, the major part of the total X-ray luminosity (about 72\%)
originates from inside the cooling radius. This corresponds to a total
radiation power of $\sim 3~ 10^{45}$ erg s$^{-1}$. So far we have no 
direct indication that this cooling rate is balanced by AGN heating in this system.
Some support for this scenario is discussed below. Assuming that the above
sketched cooling core scenario applies and that the observed cooling power
inside the cooling radius is balanced by the energy output from the AGN we 
can calculate further interesting system parameters. If the radiation power is
replenished by accretion power from the AGN and if we assume an energy
return efficiency, $\eta = 0.1$ from accretion onto the AGN black hole
we can imply an accretion rate of the order of $0.5$ M$_{\odot}$ yr$^{-1}$.
Thus, in this mode the central black hole can gain a considerable mass
of the order of the most massive black holes known over cosmic times.

Therefore it is very interesting to see if this scenario actually applies to
RXCJ1504-0248. So far the observational evidence is far less detailed as for
the best observed nearby cooling core clusters and the implications indicated
above remain very speculative. But the few additional features known, point in the
right direction. The central dominant galaxy is known to harbor a radio source
with a brightness of 62 mJy at 1.4 GHz (Bauer et al. 2000) and thus it presumably 
contains a massive black hole. The radio source image obtained from the NVSS survey
is, however, unresolved and featureless. A zoom into the central region of the
CHANDRA image (Fig. 9) shows indications of an asymmetric distortion which could be due
to the interaction of the AGN jets with the intracluster medium. Better 
photon statistics is needed to draw a more firm conclusion. Therefore 
deeper X-ray observations have been scheduled for CHANDRA and a detailed VLA radio study
has been proposed for this cluster.           

The central dominant galaxy is extremely large. The Gunn r magnitude of 16.4 
determined in the Las Campanas redshift survey (Shechtman et al. 1996) translates
into an absolute magnitude of $M_r \sim -24$ which corresponds to about $3 \times 10^{11}$
L$_{\odot}$ in this band. It places this galaxy in the upper few percent
of the luminosity distribution of cluster central galaxies as found for example 
by comparison to the survey by Lauer and Postman (1994).

The optical spectrum of the central galaxy, shown in Fig. 10, shows narrow,
low excitation emission lines, notably strong [O II], weaker [O III], 
and bright H$\alpha$/[NII] lines, similar to what has been observed for many
massive cooling flows (e.g. Hu, Cowie \& Wang 1985; Johnstone, Fabian \& Nulsen
1987, Heckman et al. 1989, Donahue, Stocke \& Gioia 1989; McNamara \& O'Connell 1993;
Crawford et al. 1999). 
This provides another hint that this cluster resembles
a scaled-up version of the known nearby cooling core clusters. 

A very similar spectrum has been observed and studied in detail in the central galaxy
of the cooling core galaxy cluster A2597 by Voit \& Donahue (1997). In the spectrum
of A2597 the [OII] line is even more dominant compared to the other lines. 
Voit \& Donahue provide a very comprehensive discussion on the origin of this 
spectrum. They conclude that hot stars constitute the best fitting source of 
ionization, but to match the spectral properties with an photoionization 
nebular model an additional heat source has to be assumed which may well be the
heating source of the cooling core. The same discussion most probably applies
also to this object, but to perform the same analysis a deeper spectroscopic
observation is required to get accurate line fluxes for more diagnostic lines.
Very similar spectra can also be found among the spectra compiled from central cluster
galaxies by Crawford et al. (1999). It is very striking that the spectra which
best match the spectrum of RXCJ1504-0248 are those from the prominent well
known cooling flow clusters, such as Z3146, A1835, A2204, and A2390. Overall these 
spectra show some variation in the degree of ionization with [OII] and [OI] lines
being more prominent compared to [OIII] in Z3146 and [OIII] being relatively more prominent
in A2390 than in the present case. But the close resemblance is very obvious.
Within the classification scheme of AGN the present observation 
features as a LINER spectrum. 

The equivalent width of the [OII] line is about $151 \pm 2.43$ {\AA} corresponding to a 
line flux of about $9 \times 10^{-14}$ erg s$^{-1}$ cm$^{-2}$. Together with the total
luminosity of the galaxy and the relation of [OII] luminosity and star formation
rate as given by Kennicutt (1992) we get formally a star formation rate of the 
order of about 50 M$_{\odot}$ yr$^{-1}$. This is not untypical for a massive cooling core
cluster (e.g. McNamara 1997).
Since there are surely other ionization 
sources present, this simple modeling is certainly an oversimplification.

\section{Conclusion} 

The CHANDRA observation reveals the X-ray source RXCJ1504-0248
as a galaxy cluster that has extreme and surprising properties in
two respects. It is found to be a very massive cluster and
the most luminous cluster known in the southern sky at
redshifts lower than $z \le 0.3$. Secondly, the cluster appears
extremely compact with a very dense central region. Thus the high
X-ray luminosity is the result of both, the large cluster mass
and the high central density.

These properties would make the cluster a cooling flow with one
of the largest mass deposition rates ever inferred. But the observation
of a radio AGN in the cluster center and the absence of low temperature
signatures in the central X-ray spectrum leads us to suspect that
the dense central ICM region is heated by the AGN like it is implied
from the X-ray observations for nearby clusters. The heating source
required in RXCJ1504-0248 needs to have a power about an order
of magnitude larger than that of the nearby well studied cooling core
clusters. This extreme energy recycling will surely make RXCJ1504-0248
a very interesting study object for many future investigations.
This X-ray source has been accepted for observations with ASTRO-E2
which should for example provide insight into the degree of turbulence
prevailing in the cooling core region.

\acknowledgments
This study was made possible by the highly advaced capabilities of the
NASA CHANDRA Observatory. We also acknowledge support for our international
collaboration by NASA under the grant G04-5142X.

\clearpage



\begin{figure}
\epsscale{.80}
\plotone{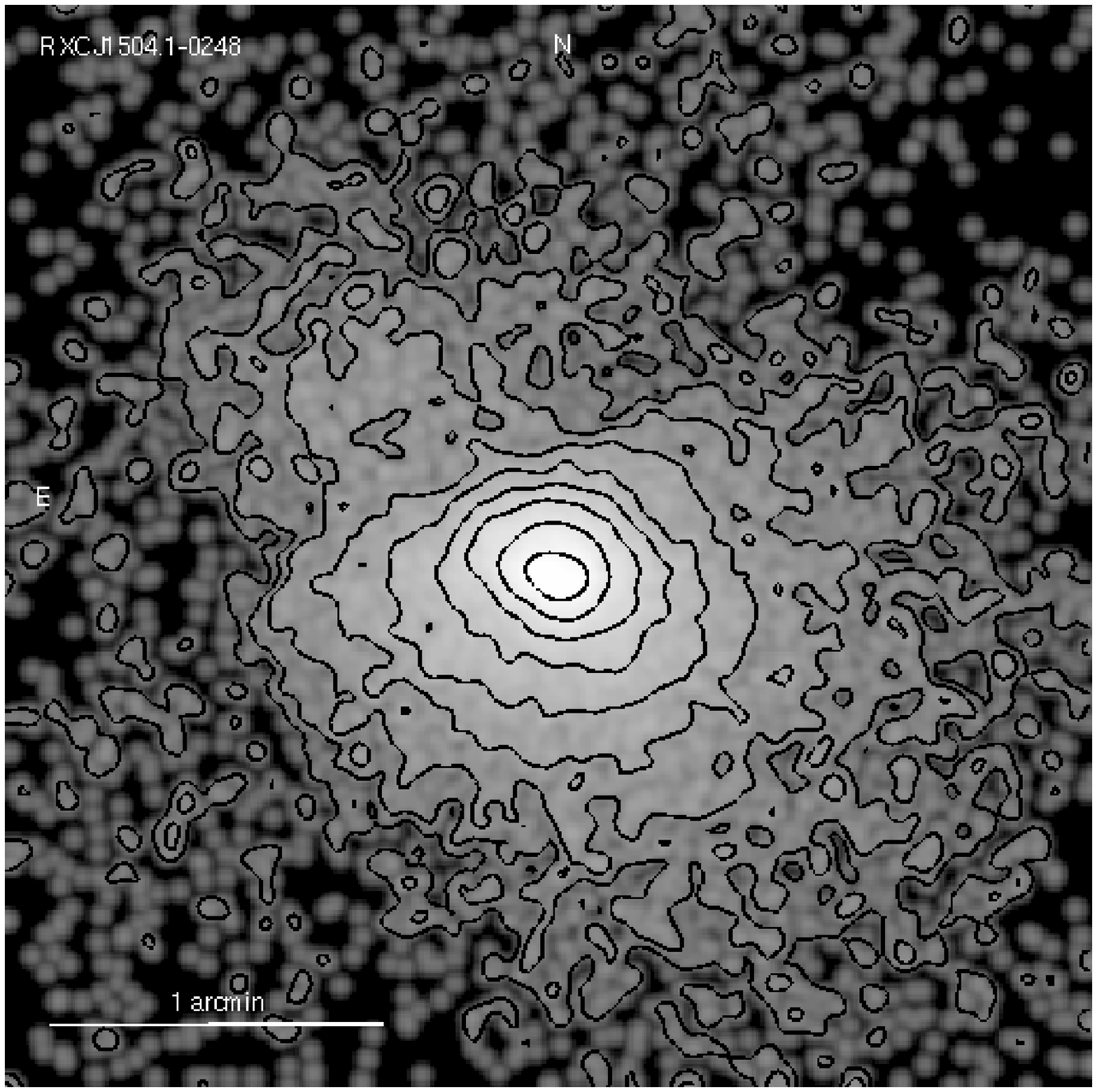}
\caption{CHANDRA ACIS I image of RXCJ1504.1-0248 in the 0.5 to 2 keV band.
The contour levels increase by factors of two. The sharp surface brightness
drop north of the cluster center with a position angle $PA \sim 115\deg$ is an 
artefact produced by the gaps in the ACIS I CCDs. North is up and East is left.
}
\end{figure}

\clearpage

\begin{figure}
\epsscale{.80}
\plotone{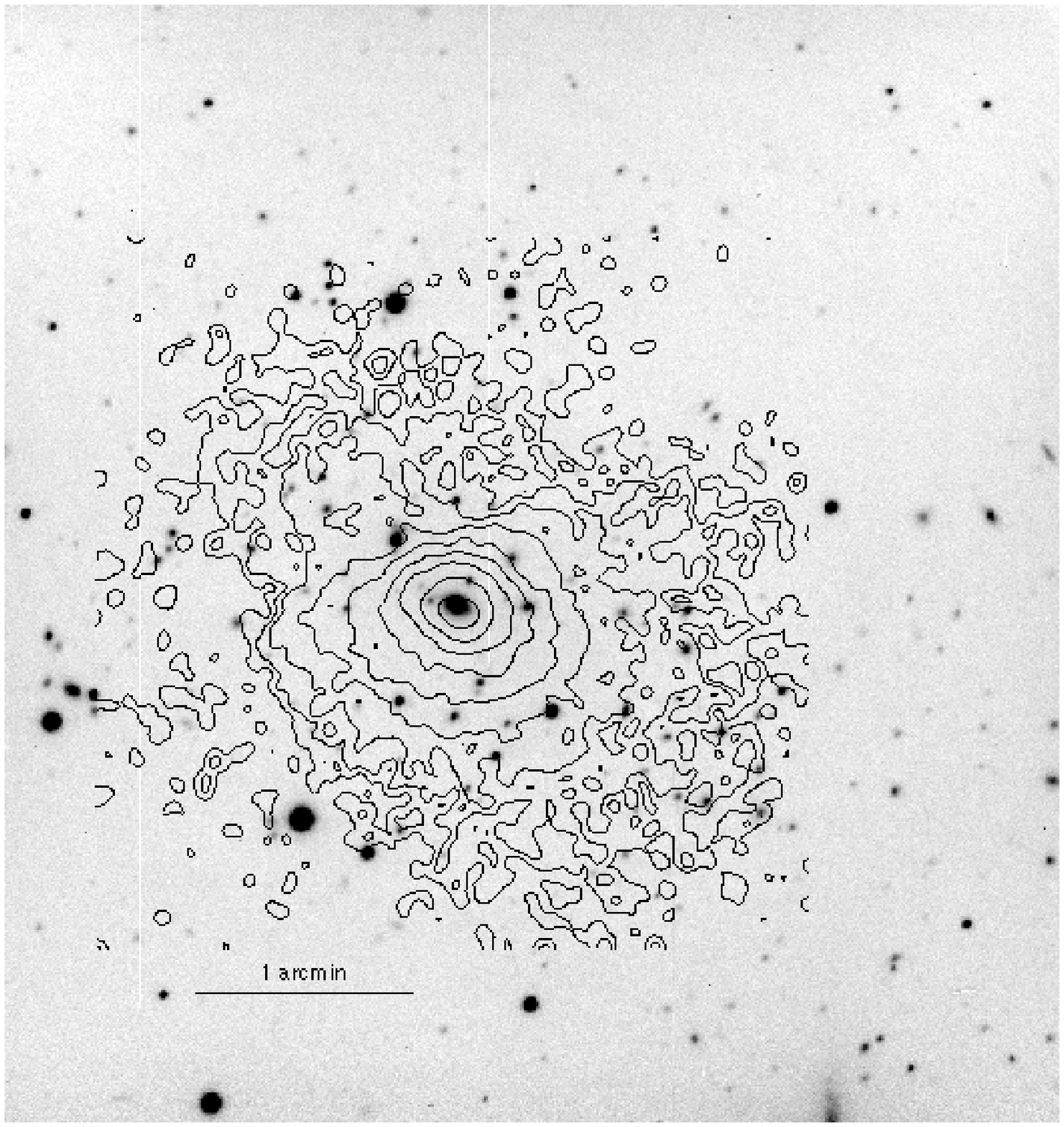}
\caption{R-band image of the cluster RXCJ1504.1-0248 taken at the
3.6m ESO telescope overlayed with the X-ray contours of Fig. 1.
 North is up and East is left.}
\end{figure}

\clearpage

\begin{figure}
\epsscale{.80}
\plotone{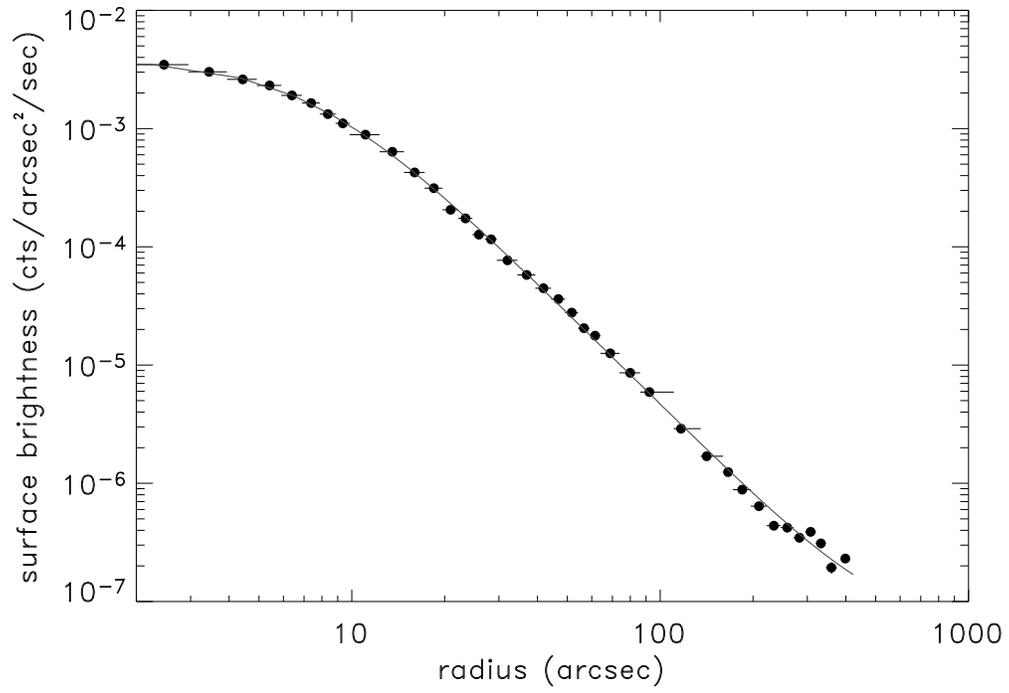}
\caption{Surface brightness profile of RXCJ1504.1-0248 in the 0.5 to 2 keV band
fit by a $\beta$-model.}
\end{figure}

\clearpage

\begin{figure}
\epsscale{.80}
\plotone{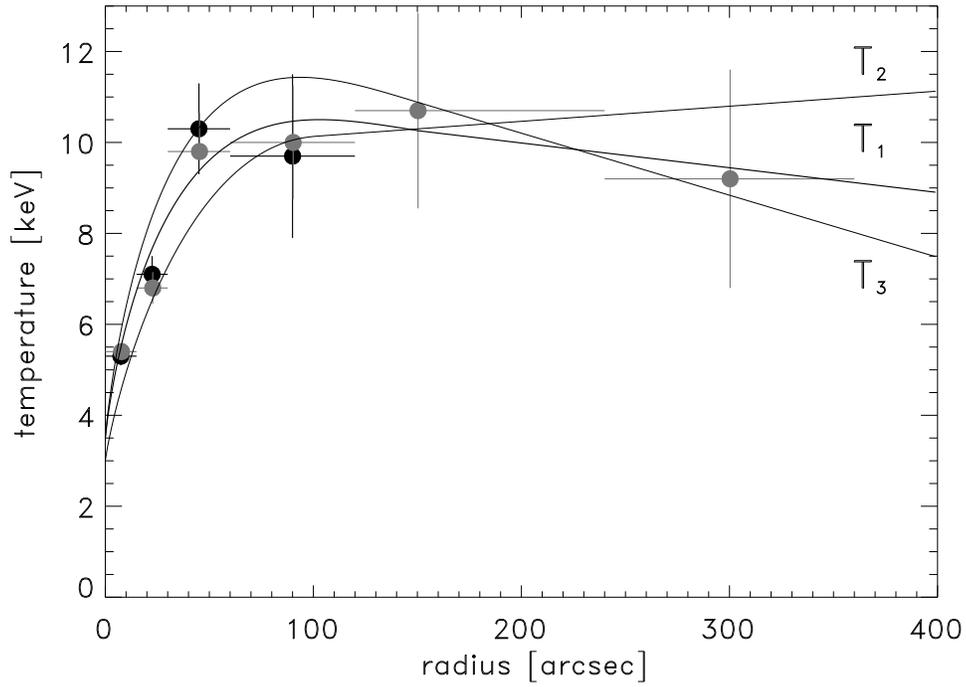}
\caption{Temperature profile of RXCJ1504.1-0248 determined from
the spectroscopic analysis of the ACIS I data in four concentric rings.
The black dots show the temperature determination with the on-target
background while the grey dots show the results of the spectral analysis
using the background field. The lines show the three analytic representations
of the temperature profile described in the text.}
\end{figure}

\clearpage

\begin{figure}
\epsscale{.80}
\plotone{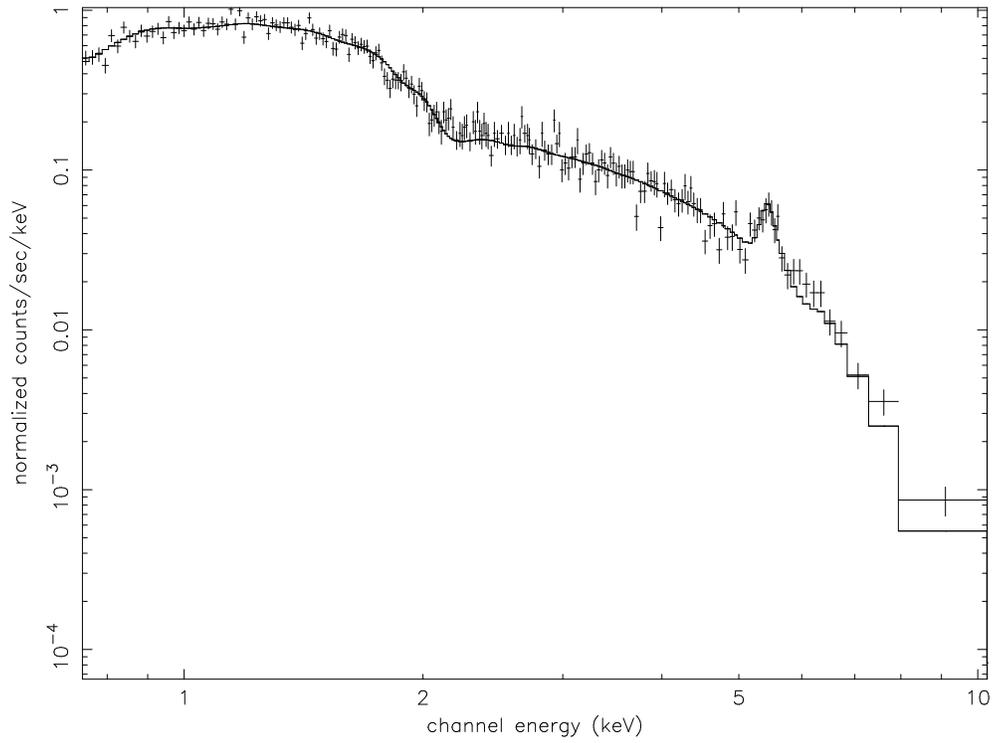}
\caption{CHANDRA ACIS I spectrum of the central 15 arcsec radius region of
the cluster RXCJ1504.1-0248. The well visible Fe line allows to confirm
the redshift and to measure the Fe abundance.}
\end{figure}

\clearpage

\begin{figure}
\epsscale{.80}
\plotone{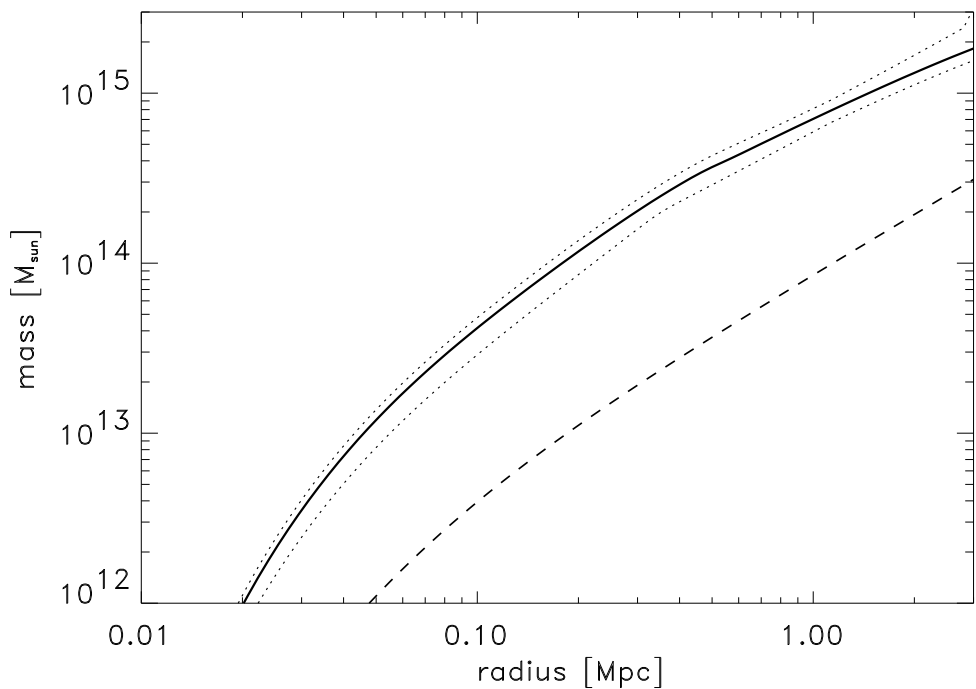}
\caption{Gravitational mass profile of RXCJ1504.1-0248 calculated for the three
bracketing gas temperature profiles adopting a $\Lambda$-cosmological model with
a Hubble constant of $H_0 = 70$ km s$^{-1}$ Mpc$^{-1}$. The two dotted lines indicate
the uncertainties determined from the local minima and maxima in the mass profiles from
the three analytic fits to the temperature profile and their scaling uncertainties
(see text). Also shown as dashed
line is the gas mass profile calculated from the $\beta$-model.}
\end{figure}

\clearpage

\begin{figure}
\epsscale{.80}
\plotone{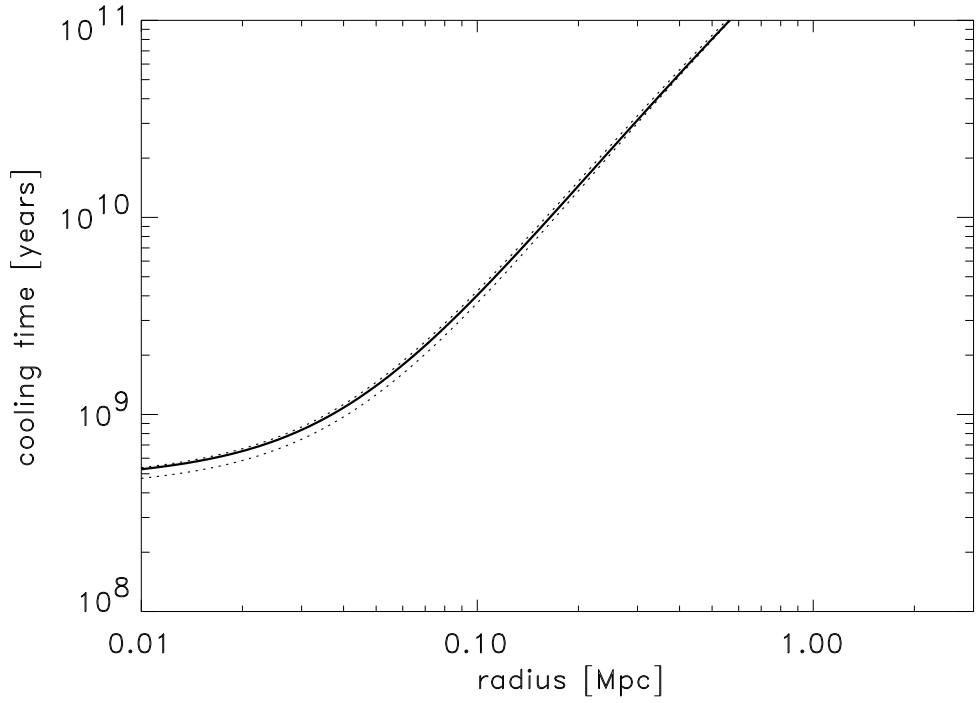}
\caption{Cooling time profile for the ICM in RXCJ1504.1-0248 calculated for the three
bracketing gas temperature profiles adopting an Einstein-de Sitter cosmological model 
with a Hubble constant of $H_0 = 50$ km s$^{-1}$ Mpc$^{-1}$.}
\end{figure}

\clearpage
\begin{figure}
\epsscale{.80}
\plotone{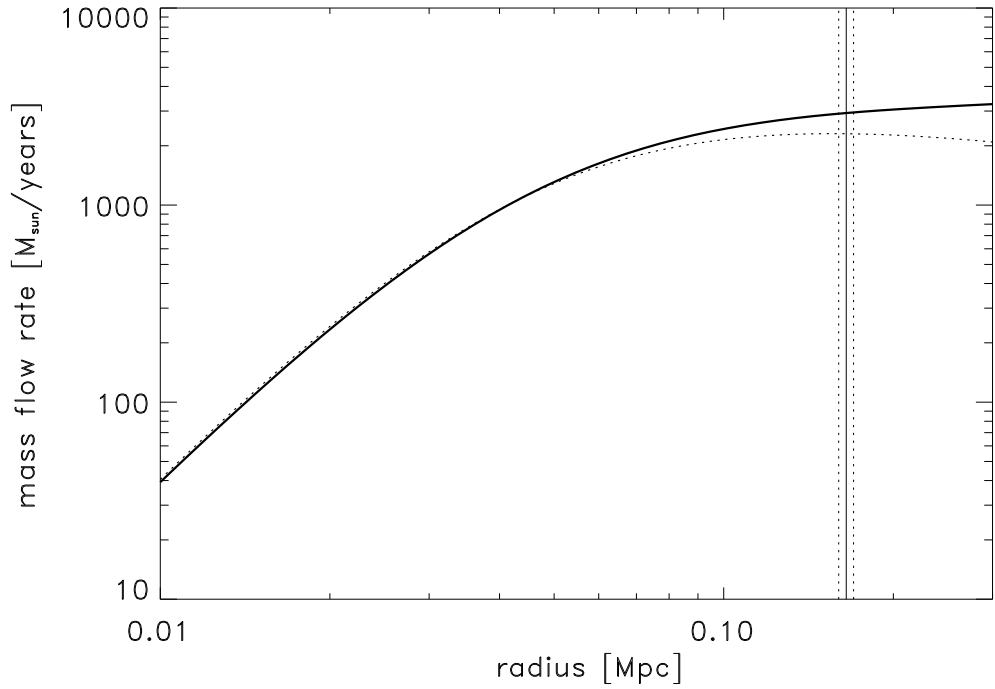}
\caption{Inferred mass flow rates assuming a conventional cooling flow model (solid line =
model A; dashed line = model B including gravitational energy gain)
as a function of radius for an Einstein-de Sitter cosmological model with
a Hubble constant of $H_0 = 50$ km s$^{-1}$ Mpc$^{-1}$. The vertical lines indicated the
radius with a cooling time of $10^{10}$ years and its uncertainty.}
\end{figure}

\clearpage

\begin{figure}
\epsscale{.80}
\plotone{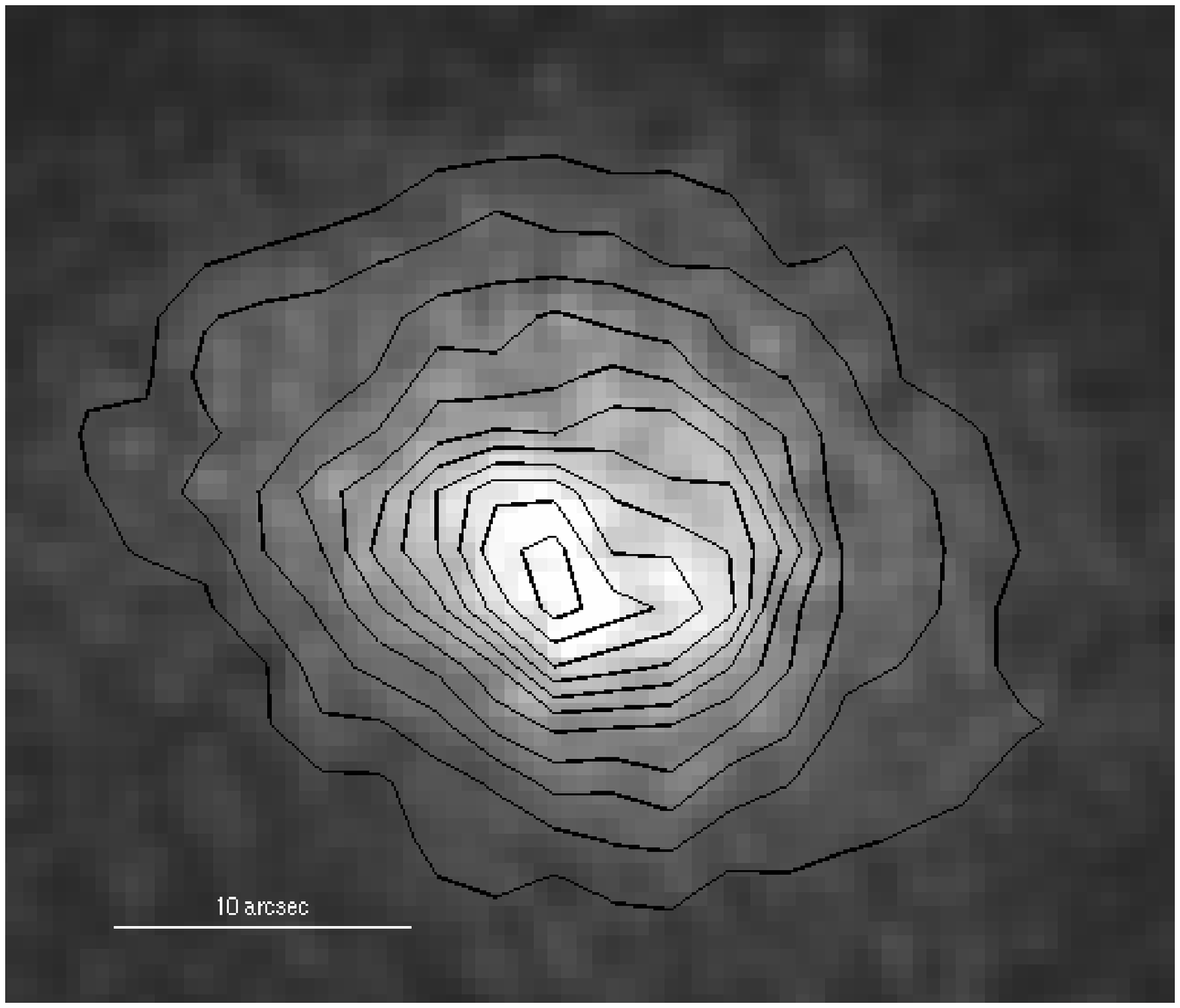}
\caption{Zoomed ACIS I image of the central region of RXCJ1504.1-0248.
Some asymmetric distortion is visible NW from the center which may be
an indication of possible interaction effects of the central AGN with the cluster ICM.
North is up and East is left.}
\end{figure}

\clearpage

\begin{figure}
\epsscale{.80}
\plotone{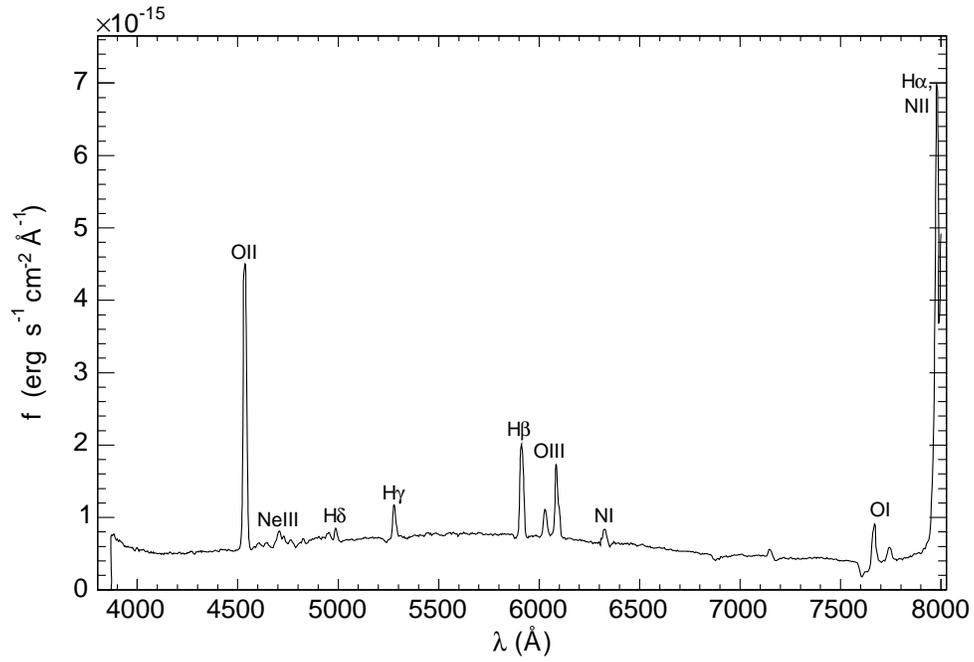}
\caption{Spectrum of the central galaxy of RXCJ1504.1-0248 obtained with
the ESO 3.6m telescope. The spectrum has LINER like characteristics quite
typical for galaxies in the centers of cooling flows. The spectrum was 
obtained with the EFOSC2 instrument with a spectral slit width of 2 arcsec
positioned through the central part of the galaxy LCRS B150131.5-023636.}
\end{figure}

\end{document}